\documentclass{ws-p8-50x6-00}
\usepackage{amsfonts}

\begin{document}
\title{A History of Feynman's Sum over Histories\\ in Quantum Mechanics} 
\author{Akira Inomata} 
\address{Department of Physics, State University of New York at Albany,\\
  Albany, New York 12222, U.S.A.}
\author{Georg Junker} 
\address{Institut f\"ur Theoretische Physik, Universit\"at Erlangen-N\"urnberg,
  Staudtstr.\ 7, D-91058 Erlangen, Germany} 

\maketitle

\abstracts{
A history of Feynman's sum over histories is presented in brief. A focus is
placed on the progress of path-integration techniques for exactly
path-integrable problems in quantum mechanics.}

\section{Classification of Path Integrals}
Exact calculations of Feynman's path integrals (defined on a time 
lattice) are mainly based on recurrence integral formulas in which the 
convolution of two functions having a common feature retains the same 
feature. Therefore, exactly soluble path integrals in quantum mechanics 
may be classified by their recurrence integral formula used in the 
calculation. According to this classification, there are three types of 
path integrals: (i) {\em Gaussian path integrals}, (ii) {\em Legendrean 
path integrals}, and (iii) {\em Besselian path integrals}. The Gaussian 
path integrals are calculated by the well-known convolution of two 
Gaussian functions which produces recurrently another Gaussian function. 
Path integrals of this type have been widely used in semiclassical 
approaches, coherent-state path integrals, applications in statistical 
physics, field theory and many others areas. The Legendrean path 
integrals are based on the convolution integral for zonal spherical 
functions (generalized Legendre functions), which are particular matrix 
elements of unitary irreducible representations of Lie groups. An 
elementary type of the Legendrean path integrals appears in the angular 
path integral in 3-dimensional polar coordinates. More generally, path 
integrals of this type are useful for  systems with certain symmetries 
of kinematical or dynamical origin. For details see the review [523]. 
The Besselian path integrals are based on Weber's integral formula for 
Bessel functions [771] or the group composition law (convolution) for 
particular unitary representations of an element of $SU(1,1)$ in a 
continuous base. Radial path integrals are of the Besselian type and may 
be associated with a certain dynamical group or spectrum generating 
algebra [523]. 
 
The books of Feynman and Hibbs [340] and of Schulman [828]  discuss 
mainly path integrals of the Gaussian type which are undoubtedly most 
important in applications. However, to understand the complete feature 
of Feynman's path integral, we cannot ignore the non-Gaussian aspects. 
Indeed, the non-Gaussian features have been found essential in studying 
exactly path-integrable systems. The history of those non-Gaussian path 
integrals began with the polar-coordinate formulation of path integrals. 

\section{Polar Coordinate Formulation and Harmonic
analysis}

In the early 1950's, Ozaki (lecture notes, Kyushu University, 1955, 
unpublished) started with a short-time action for a free particle 
written in cartesian coordinates and transformed it to the polar form. 
Then he performed angular path integration but failed in carrying out 
explicitly the radial path integral. In 1964, Edwards and Gulyaev [291] 
formulated the free particle propagator in polar coordinates in a way 
similar to that of Ozaki. They have completed the radial path 
integration for the free particle. In 1969, Peak and Inomata [771] 
calculated explicitly the radial path integral for the harmonic 
oscillator (in an inverse-square potential). This opened the direction 
of the Besselian path integral which became an important base for the 
study of exactly path-integrable systems [528]. 

In addition it became clear that group theoretical methods, most notably 
harmonic analysis, are not only elegant but also powerful tools in path 
integration. In 1968 Schulman [826] discussed spin with the path 
integral of a rigid rotor in terms of $SU(2)$ representations. In 1970 
Dowker [262] discussed exactness of semiclassical results on compact 
groups such as $SU(n)$. Marinov and Terentyev [679], in 1979, used 
harmonic analysis for path integration on $SO(n+1)/SO(n)$. In 1987 
B\"ohm and Junker [104] generalized this to formulate path integrals on 
symmetric spaces including $SU(1,1)$ and $SO(n,1)/SO(n)$, and solved 
chiefly Legendrean-type path integrals by harmonic analysis. In 1991 a 
group theoretical treatment of the radial path integrals was made on the 
basis of $SU(1,1)$ by the present authors [523]. 

\section{The Hydrogen-Atom Problem}
Before 1979 the list of exactly path-integrable systems was very short. 
It is a curious fact that Feynman's path integral could not reproduce 
the exact solution for the hydrogen atom which once symbolized the 
success of Schr\"odinger's wave mechanics. If Feynman's approach is 
equivalent to Schr\"odinger's, then it should be able to solve the very 
standard problem in quantum mechanics. Since the path integral involving 
the Coulomb potential was not directly integrable, Gutzwiller [749] 
treated it semiclassically on the basis of  his famous trace formula, 
yielding the exact energy spectrum, whereas Goovaerts and Devreese [408] 
made a perturbation calculation and obtained the exact $s$-wave 
spectrum. 

In 1979, however, Duru and Kleinert [279] made a breakthrough in solving 
the hydrogen problem by path integration without approximations. 
They applied the Kustaanheimo-Stiefel (KS) space-time transformation to 
the path integral. The KS transformation, which consists of a coordinate 
map from ${\mathbb R}^{3}$ to ${\mathbb R}^{4}$ and a path-dependent 
time transformation, converted the path integral for the hydrogen atom 
into the exactly soluble path integral for an isotropic harmonic 
oscillator in four dimensions. 

\section{Other Exactly Path-Integrable Systems}
Although the KS transformation is limited in application, its success in 
the hydrogen atom suggested that (i) a local time transformation may be 
used in combination with a nonlinear coordinate transformation in order 
to reduce a non-soluble path integral to a soluble path integral; and 
(ii) the dimension of the configuration space for path integration may 
be extended so as to treat a dynamical symmetry of the system in 
question as a kinematical one. By these techniques, 
nearly all the exactly soluble problems by Schr\"odinger's equation 
became path-integrable. The local time transformation was taken only as 
a formal trick in the beginning, but its use was justified 
for the Wiener-type path integrals (Feynman path integrals in Euclidean 
time) by Blanchard and Sirugue [97], Young and DeWitt-Morette [943], 
Castrigiano and St\"ark [152], and Fischer, Leschke and M\"uller 
[343,344]. 

Finally we wish to remark that there are two types of time 
transformations applicable to path integration. One is integrable 
(globally meaningful), whereas the other is nonintegrable (only locally 
meaningful). The KS time transformation is of the latter case. A 
remarkable example involving an integrable time transformation is the 
one considered by Alfaro, Fubini and Furlen and by Jackiw, 
which converts a harmonic oscillator into a free particle [128,552 and
references therein].  

\section*{References}
There are a number of textbooks on Feynman's path integral. The most 
recent one is C.\ Grosche and F.\ Steiner, "Handbook of Feynman Path 
Integrals", (Springer, Berlin, 1998), which has an extensive list of 
references (pp. 368-423) and textbooks (pp.21-22). Specifically the 
book [528] in the list on page 21 discusses in detail several techniques 
used for exact path integration in quantum mechanics. In the present 
article, for convenience, we adopt the reference numbers of the book of 
Grosche and Steiner; the numbers given in square brackets in the text 
refer to the corresponding reference numbers. 
\end{document}